\documentclass[twocolumn]{IEEEtran}

\usepackage{cite}
\usepackage{gensymb}
\usepackage{amsmath,amssymb,amsfonts}
\usepackage{bm}
\usepackage{amsthm}
\usepackage{textcomp}

\usepackage{algorithm}
\usepackage{algpseudocode}

\usepackage{soul} 

\usepackage{graphicx}
\usepackage{textcomp}
\usepackage{xcolor}
\usepackage{comment}
\usepackage{svg,balance}
\usepackage{subcaption}
\usepackage[margin=20mm,top=20mm,bottom=20mm]{geometry}
\usepackage[normalem]{ulem} 
\usepackage{cancel}
\usepackage{mathtools, cuted}

\usepackage{tikz}
\usetikzlibrary{shapes,arrows,positioning}
\usetikzlibrary{3d}

\usepackage[nodisplayskipstretch]{setspace}
  {\proof}{\proofend}

\newcommand{\qh}{{\bf h}}

\newcommand{\qp}{{\bf p}}

\DeclareMathOperator*{\argmin}{arg\,min}

\newcommand{\MRT}{\mathrm{MRT}}

\newcommand{\ps}{{\qp_{s}}} 
\newcommand{\pmr}{{\qp_{m}}} 

\newcommand{\gsm}{{\textbf{g}_{s,m}}} 
\newcommand{\gskt}{{\textbf{g}^T_{s,k}}} 

\newcommand{\gsmc}{{\textbf{g}^*_{s,m}}} 

\newcommand{\kmrt}{{\kappa^{\MRT}_{s,m}}}
\newcommand{\kmrtp}{{\kappa^{\MRT}_{s',m}}}

\newcommand{\wsm}{{\textbf{w}_{s,m}}} 

\newcommand{\as}{{a_{s}}} 
\newcommand{\asi}{{a^i_{s}}} 
\newcommand{\ats}{{\tilde{a}_{s}}} 

\newcommand{\asp}{{a_{s'}}} 

\newcommand{\nx}{{N_{s,x}}} 
\newcommand{\ny}{{N_{s,y}}} 
\newcommand{\ns}{{N_s}} 

\newcommand{\sus}{{\sum\nolimits^{S}_{s=1}}}
\newcommand{\summ}{{\sum\nolimits^{M}_{m=1}}}

\newcommand{\susp}{{\sum^{S}_{s'=1}}}

\newcommand{\IEM}{{{I}^E_{M}(\os,\!\as)}} 

\newcommand{\IEMt}{{{I}^E_{M}(\os,\!\ats)}} 

\newcommand{\PC}{{P_{c}(\os,\!\as)}} 
\newcommand{\PCt}{{P_{c}(\os,\!\ats)}} 

\newcommand{\gamEff}{{\Gamma(\os,\!\as)}} 
\newcommand{\gamEfft}{{\Gamma(\os,\!\ats)}} 

\newcommand{\os}{{\Omega_{s,m}}}

\newcommand{\ots}{{\tilde{\Omega}_{s,m}}}
\newcommand{\osim}{{\tilde{\Omega}^{i-1}_{s,m}}}
\newcommand{\osp}{{\Omega_{s',m}}}

\newcommand{\gos}{{g_s(\ots)}}
\newcommand{\gios}{{g^i_s(\ots)}}

\newcommand{\ME}{{\mathbb{E}}} 

\def\BibTeX{{\rm B\kern-.05em{\sc i\kern-.025em b}\kern-.08em
    T\kern-.1667em\lower.7ex\hbox{E}\kern-.125emX}}
    
\title{Near-Field Energy Harvesting Using XL-MIMO Over Non-Stationary Channels}
\author{Muhammad Zeeshan Mumtaz,~\IEEEmembership{Student Member,~IEEE,} Mohammadali Mohammadi,~\IEEEmembership{Senior Member,~IEEE,} 
Hien Quoc Ngo,~\IEEEmembership{Fellow,~IEEE,}   and  Michail Matthaiou,~\IEEEmembership{Fellow,~IEEE}\\
\thanks{

This work was supported by the U.K. Engineering and Physical Sciences Research Council (EPSRC) (grants No. EP/X04047X/1 and EP/X040569/1); the U.K. Research and Innovation Future
Leaders Fellowships under Grant MR/X010635/1; a research grant from the Department for the Economy Northern Ireland under the US-Ireland R\&D Partnership Programme; and the European
Research Council (ERC) under the European Union’s Horizon 2020 research
and innovation programme (grant agreement No. 101001331).}
\thanks{The authors are with the Centre for Wireless Innovation (CWI), Queen's University Belfast, BT3 9DT Belfast, U.K. (email:\{mmumtaz01, m.mohammadi, hien.ngo, m.matthaiou\}@qub.ac.uk).}
\thanks{M. Z. Mumtaz is also with the College of Aeronautical Engineering, National University of Sciences \& Technology (NUST), Pakistan (email: zmumtaz@cae.nust.edu.pk).}
}
\date{}

\begin{document}

\maketitle
\begin{abstract}
    This paper explores the  maximization of the harvested power efficiency (HPE) in a modular extremely large multiple-input multiple-output (XL-MIMO) system, which supports energy harvesting (EH) for  near-field users. These users are located in spatially distinct visibility regions (VRs) with non-stationary channel characteristics. We propose to determine which sub-arrays are switched on or off as well the power control coefficients at the sub-arrays to maximize the HPE. The design can be processed via a multi-tier joint optimization framework based on fractional programming. The numerical results showcase that the HPE performance of the proposed algorithm is nearly optimal, comparable to that of exhaustive search. As a matter of fact, it achieves up to a $120\%$ gain over the benchmark scheme which uses the entire XL-MIMO array with equal power allocation (PA) across sub-arrays, while significantly reducing the computational time.
\end{abstract}
\begin{IEEEkeywords}
Energy harvesting (EH), extremely large multiple-input multiple-output (XL-MIMO), harvested power efficiency (HPE).
\end{IEEEkeywords}

\vspace{-1.5em}
\section{Introduction}

Extremely large multiple-input multiple-output (XL-MIMO) antenna arrays are one of the key technologies for future 6G mobile networks. 
In addition to the hardware structure differences with conventional massive  MIMO (mMIMO), XL-MIMO entails several new features and opportunities that are not present in conventional mMIMO. First, the substantial increase in the array aperture has motivated the modeling of the electromagnetic (EM) radiation field by near-field spherical waves, in contrast to the planar-wave-based radiation model of mMIMO~\cite{Haiquan}. Spherical wavefronts stimulate the generation of radiation patterns that focus the beam at a specific location, referred to as beamfocusing~\cite{Haiyang}.
Focused beams facilitate the interference mitigation in multi-user communications and have been envisioned as a method to enable efficient wireless power transfer (WPT) with minimal energy dissipation~\cite{Haiyang}. Secondly, the spatial non-stationarity (SnS) of the channel, where only part of the array is visible to the users (known as the visibility region, VR)~\cite{cui2}, must be also considered during the system design.


Recent near-field WPT studies promote energy focusing for efficient EH. For example, \cite{Haiyang} used dynamic metasurface antennas to direct power and reduce energy leakage, while \cite{Zhang:JSAC:2024, Zhang:IoT:2024} explored joint beam scheduling, power allocation, and hybrid beamforming for SWIPT. However, these works overlook the SnS effects, like spherical propagation and blockage. By leveraging SnS-induced VRs, where users mainly receive power from specific sub-arrays, this letter investigates strategic sub-array activation to enhance the harvested power efficiency (HPE).

The main contributions of the paper are two-fold: 1) we develop an optimization framework to maximize the HPE of WPT-based XL-MIMO, subject to a joint PA and sub-array activation (SA). By leveraging the multi-layered optimization approach from~\cite{yuhan}, we split the primary complicated objective into two sub-problems. 2) For the PA sub-problem, the Douglas-Rachford (DR) splitting method~\cite{pontus} is applied alongside the Dinkelbach's transform based fractional programming~\cite{Shen}. The SA is parameterized using a surrogate auxiliary function of the optimized power coefficients, similar to the model proposed in~\cite{koziel}. Our simulation results demonstrate that the HPE performance of the proposed algorithm closely matches that of the optimized PA achieved through exhaustive search (ES) across the favorable active sub-arrays, while significantly reducing the computational time. 

\emph{Notations:} We use bold lower case letters to denote vectors. The superscripts $\!(\cdot)^{\rm{*}}\!$ and $\!(\cdot)^{\rm{T}}\!$ denote the conjugate and transpose of a matrix, respectively; $\| \cdot \|$ returns the norm of a vector. Finally, $\mathbb{E}\{\cdot\}$ denotes the statistical expectation.
\section{System Model}
\vspace{-0.2em}
We consider a wireless communication system, where a modular XL-MIMO array supports the  EH functionality  to $M$ users located in $V$ distinct spatial regions, where $V\leq M$.  This huge array consists of $S$ sub-arrays (modules) which are two dimensional uniform planar arrays (2-D UPAs). Each of these sub-arrays is equipped with $\ns$ antenna elements, consisting of  $\nx$ elements in the $x$-axis and $\ny$ elements in the $y$-axis.
The user devices are located within the radiative near-field region of the overall array, where the EM waves manifest spherical wavefront nature with finite beam depth. This region extends up to the one-tenth of Fraunhofer array distance $d_{f}$~\cite{Ramezani2024}. This physical parameter is defined as $d_{f}=2 D^2 (S \ns)/\lambda$, where $D$ is the largest dimension of each antenna element and $\lambda$ is the wavelength of the carrier frequency. The free-space near-field radiating channel $\gsm \in \mathbb{C}^{\ns \times 1}$ from  sub-array $s$  to  user $m$ at location $\pmr= (x_m,y_m,z_m)$, which stems from the cosine pattern model \cite{Haiquan,Haiyang}, is given as
\vspace{-0.0em}
\begin{equation}\label{eq:channel}
    \gsm = \dfrac{\lambda}{4 \pi \lVert \pmr -\ps \rVert}\sqrt{F(\theta_{s,m})} \qh_{s,m} ,
\end{equation}
where
\vspace{0.5em}
\begin{align*}
    \qh_{s,m}\!\!= \!\!\Big[e^{\!-j k \lVert \pmr\! -\! \qp_{s_{1,1}} \! \rVert},\!e^{\!-j k \lVert \pmr\! - \!\qp_{s_{1,2}}\!  \rVert}\!,\!\hdots\!,\! 
e^{\!-j k \lVert \pmr \!-\! \qp_{s_{N_x,N_y}}\!  \rVert}\Big]\!,
\end{align*}
represents the channel vector from sub-array $s$, whose antenna elements are positioned at the locations $\ps\in\{(x_{s_{1\!,1}},\!y_{s_{1\!,1}},\!0),\! (x_{s_{1\!,2}},\!y_{s_{1\!,2}},\!0),\! \hdots\!, \!(x_{s_{N_x\!,N_y}},\!y_{s_{{N_x\!,N_y}}},\!0)\}$; $k\!=\!2 \pi/\lambda$ is the wave number; $e^{-j k \lVert \pmr -\ps \rVert}$ denotes the phase attributed to the distance covered by the EM wave from $\ps$ to $\pmr$; $F(\theta_{s,m})$ is the cosine radiation profile of each antenna element for the boresight gain $b$, defined as
\vspace{-0.0em}
\begin{align}
    F(\theta_{s,m})\!&=\! 
    \begin{cases}
        2 (b+1) \,\text{cos}^{b}(\theta_{s,m}) \quad  \theta_{s,m} \in [0,\pi/2],\\
        0 \hspace{3cm} \text{otherwise}.
    \end{cases}
\end{align}

We consider a fully-digital antenna array configuration with dedicated radio-frequency (RF) chains for each antenna element. Furthermore, we design the maximum ratio transmission (MRT) precoding for  user $m$ from the sub-array $s$ as $\wsm= \kmrt \gsmc$, where $\kmrt \!\!=\! 1/\lVert\gsm\rVert$.  Note that MRT precoding has been established as the optimal precoding method for WPT, especially with a large number of antennas~\cite{cite:MRT_for_HE:Almradi}. With regards to the SA, the transmitted signal by the sub-array $s$ is expressed as 
\vspace{-0.1em}
\begin{align}~\label{eq:xs}
   \textbf{x}^E_s= a_s\summ   \sqrt{\os} \wsm e_m,
\end{align}
where $\as$ represents the SA binary variable, defined as
\vspace{-0.0em}
\begin{align}
    \as &= 
    \begin{cases}
        1 \quad \quad \text{if sub-array $s$ is switched on},\\
        0 \quad \quad \text{if sub-array $s$ is switched off}.
    \end{cases}
\end{align}
Moreover in~\eqref{eq:xs}, $e_m$ is the normalized zero-mean energy symbol, satisfying the condition $\ME\{\lvert e_m \rvert^2\}=1$, while $\os$ denotes the PA coefficient between sub-array $s$ and user $m$, chosen to satisfy the power constraint at each sub-array as
\vspace{-0.2em}
\begin{equation}
    \mathbb{E}\left\{\lVert \textbf{x}^E_s \rVert^2 \right\}= \sum\nolimits^{M}_{m=1} \os \leq P_s,
\end{equation}
where $P_s= \ns P_{et}$ is the maximum sub-array power, while $P_{et}$ is the maximum transmit power of each antenna element. As a result, the combined transmit power of all sub-arrays is $P_t = S P_s$. Moreover, the overall power consumed by the whole XL-MIMO system  can be modeled as~\cite{Asaad},
\begin{equation}\label{eq:power_consumed}
    \!\!\PC\!\! =\! \!\!\sum^{S}_{s=1} a_s \bigg(\frac{1}{\varsigma}\! \sum^{M}_{m=1}\! \os \! +\! 2 P_{\mathrm{syn}}\!+ \!N_s P_{\mathrm{ct}}\!\bigg)\!\!+\! M \! P_{\mathrm{cr}},
\end{equation}
where $\varsigma$ is the efficiency of the power amplifier, $P_{\mathrm{syn}}$ is the power consumed by the frequency synthesizer of each sub-array, $P_{\mathrm{ct}}$ is the circuit power consumed by each antenna's RF chain, while $P_{\mathrm{cr}}$ is the circuit power at each user's receiver RF chain. 

During the EH phase, the received signal at the $k$-th user can be expressed as
\begin{equation}
    r_{k} \!\!=\! \sum\nolimits^{M}_{m=1} \sum\nolimits^{S}_{s=1} \!\!\! \as \kmrt  \sqrt{\os} \gskt \gsmc e_m\!+\! n_{s,m},
\end{equation}
where $n_{s,m}$ denotes the noise power at user $m$. The total received RF power at the EH circuits of all the network users, when ignoring the noise power, is \cite{Zhang:IoT:2024}
\begin{align}\label{eq:harvested_power}
    \IEM &= \sum^{M}_{k=1} \sum^{M}_{m=1} \left\lvert \sus \, \as \kmrt  \sqrt{\os} \gskt \gsmc \right\rvert^2\nonumber\\
    &\hspace{-1cm}= \sum^{M}_{k=1}\sum^{M}_{m=1} \sum^{S}_{s=1} \susp \as \asp \kmrt \kmrtp \sqrt{\os \osp}\nonumber \\
    & \hspace{-0.6cm} \times (\gskt \gsmc \textbf{g}^T_{s',m}\textbf{g}^*_{s',k} ). 
\end{align}

Specifically, for the downlink EH phase, the HPE can be interpreted as the ratio  of the sum harvested power at all users to the total power consumed by the XL-MIMO system, including the overall transmitted power and the circuitry power dissipation. Therefore,  the HPE of the system can be mathematically represented as
\begin{equation}
\label{eq:EE}
    \gamEff = \dfrac{\IEM}{\PC}.
\end{equation}

\vspace{-0.5em}
\section{Optimization Problem Formulation}
In this section, we propose a joint optimization framework for the considered XL-MIMO system that determines which sub-arrays are switched on/off ($a_s=1$ or 0) and selects the power coefficients $\os$ to maximize the HPE.\footnote{ The prospective practical applications of our proposed joint PA-SA optimization framework for XL-MIMO systems include the concentrated energy beamfocusing and VR association for SA by exploiting the SnS effects in near-field communication scenarios.} Hence, the optimization problem can be formulated as:
\begin{subequations}~\label{eq:optimization_1}
    \begin{align}
        \mathcal{P}_1: &\max_{\big\{\os,\as\big\}} \,\,\gamEff\\      
        &\hspace{0em} \text{s.t.} \hspace{1em} \sus \as \summ \os  \leq P_t,\label{eq:optimization_1_C1}\\
        &\hspace{2em} \summ \os \leq P_{s},\,\, \forall 
        \,\, s=1, \hdots, S, \label{eq:optimization_1_C2}\\
        &\hspace{2em}\os\geq 0, \,\, \forall 
        \,\, s=1, \hdots, S, m=1, \ldots, M,\\
        &\hspace{2em} \as \in \{0,1\}, \,\, \forall 
        \,\, s=1, \ldots, S.
    \end{align}
\end{subequations}

In $\mathcal{P}_1$, the power constraint \eqref{eq:optimization_1_C1} guarantees that the sum transmitted power of the active sub-arrays does not exceed the combined transmission power available at the overall XL-MIMO system, while \eqref{eq:optimization_1_C2} represents the power constraint at each individual sub-array.

The optimization problem $\mathcal{P}_1$ is a complicated mixed-integer non-convex problem, due to the binary variables involved. To address this issue, we conceive a multi-tier iterative optimization scheme, in which we parameterize the problem \eqref{eq:optimization_1} by decoupling the SA variables from the sub-array power coefficients. We will first focus on the optimization process for PA based sub-problem, and then discuss the characterization approach for the SA variables.

\vspace{-0.8em}
\subsection{Power Allocation Update (PA procedure)}
\label{subsec:poweralloc}
For a particular SA choice ($\ats$) as specified in the next section, it can be ascertained that $\gamEfft$ is a quasi-concave ratio of a concave function $\IEMt$ and an affine function $\PCt$, which can be reformulated using Dinkelbach's transform.
This mathematical transformation converts the ratio of these two functions into their linear combination $\phi(\os,\ats,\lambda)$ with a new auxiliary variable $\lambda$. This  function has concave nature over the variable $\os$ for a fixed $\lambda$ and parameterized activation variable $\ats$, which is given as
\begin{equation}\label{eq:Dinkelbach}
     \phi(\os,\ats,\lambda)=\IEMt- \lambda \PCt.
\end{equation}

The conic representation of the transformed function with inversion operation in \eqref{eq:Dinkelbach} can be formulated for obtaining an update for the current iteration $t$ of the PA routine:
\vspace{-0.1em}
\begin{align}~\label{eq:conic}
    \Omega^{t}_{s,m}=\argmin\nolimits_{\big\{\os \big\}} \,\,\big(-\phi(\Omega^{t-1}_{s,m},\tilde{\as},\lambda_{t})\big).
\end{align}

To this point, we can observe that the transformed objective is the sum of two monotone functions, which can be maximized using the DR splitting method. From any initial $z^0 \in \mathbb{R}$ for the $u$ sub-iteration within the iteration $t$ of the PA procedure, the following DR splitting step is carried out:
\vspace{-1em}
\begin{subequations}~\label{eq:DR_method}
     \begin{align}
        x^{u+1}&= \text{prox}_{\{\lambda_t  P_{c} (\os,\tilde{a}_s\}}(z^{u}),\\
        y^{u+1}&= \text{prox}_{\{-I^{E}_{M} (\os,\tilde{a}_s)\}}(2x^{u+1}-z^{u}),\\
        z^{u+1}&=z^{u}+(y^{u+1}-x^{u+1}),
    \end{align}
\end{subequations}
where the proximal operator is defined as:
\begin{align}
    \text{prox}_{\Phi}(z)= \argmin_{x}\Big \{\Phi(x)+\dfrac{1}{2}\lVert x-z\rVert^2\Big\}.
\end{align}

This operator $\text{prox}_{\Phi}(z)$ essentially minimizes a convex function $\Phi(\cdot)$ with a quadratic penalty. 
If a solution to the problem \eqref{eq:optimization_1} exists, then the update variables in \eqref{eq:DR_method} satisfy $\lVert y^{u}-x^{u}\rVert \rightarrow 0$, $y^{u} \rightarrow y^*$, and $z^{u}\rightarrow z^*$, where $y^*\in \mathbb{R}$ is a solution. After the completion of the DR splitting procedure, the parameter $\lambda_t$ is updated for the next iteration:
\begin{align}~\label{eq:lambdat1}
     \lambda_{t+1}= \dfrac{I^{E}_{M} (\Omega^{t}_{s,m},\tilde{a}_s)} {P_{c} (\Omega^{t}_{s,m},\tilde{a}_s) }.
\end{align}
The PA procedure is culminated upon satisfaction of the condition $\lvert \IEMt - \lambda \PCt \rvert \leq \epsilon$ for $\epsilon>0$, providing an estimate $\ots$ for the parameterized $\ats$\footnote{The PA procedure satisfies the convergence condition of both Dinkelbach’s transform ($\lim_{t\rightarrow \infty} \phi(\Omega^{t}_{s,m},\tilde{\as},\lambda_{t}) = 0, \, \lim_{t \rightarrow \infty} \Omega^{t}_{s,m} = \Omega^{*}_{s,m}$) and DR splitting ($\lVert y^{u}-x^{u}\rVert \rightarrow 0$, $y^{u} \rightarrow y^*$, and $z^{u}\rightarrow z^*$, where $y^*\in \mathbb{R}$, as the update variable approaches optimality $x^{u} \rightarrow x^{*}$ with sub-iteration $u \rightarrow \infty$. }.
\begin{algorithm}[t]
\caption{Joint Optimization of Sub-array Activation and Power Allocation}\label{alg:opt_process}
\begin{algorithmic}[1]

\State \textbf{Initialize:} Set iteration indices $i = 0$ and $t = 0$, convergence thresholds $\delta\!>\!0$ and $\epsilon\!>\!0$, and initial $\lambda_0$.
\State Initial iterates: $\Omega^{(0)}_{s,m}=P_s/M$, $a^{0}_s=\{1\}^S$
\State \textbf{SA procedure:}
\Repeat

    \State Compute the auxiliary function $\gios$ using~\eqref{eq:gios}
    \State Update the SA variable $\asi$:
    \begin{align}
        \asi &= 
        \begin{cases}
            1 \quad \quad \gios \geq \ME\{\gios\},\nonumber\\
            0 \quad \quad \gios < \ME\{\gios\}\nonumber.
        \end{cases}
    \end{align}
    \State Scale the activation variable $\ats$ using~\eqref{eq:parameterized}
    \State Formulate the parameterized optimization problem:
    \begin{equation}
        \max\nolimits_{\{\os\}} \,\,\Gamma(\os,\tilde{a}_s^{i}) \nonumber
    \end{equation}
    \State\textbf{PA procedure:}
    \Repeat
        \State Apply Dinkelbach's transform for iteration $t$:
        \begin{equation}
             \phi(\Omega^{t-1}_{s,m},\tilde{a}_s^{i},\lambda_{t})=I^E_M(\Omega^{t-1}_{s,m},\tilde{a}_s^{i})- \lambda_{t} P_c(\Omega^{t-1}_{s,m},\tilde{a}_s^{i}).\nonumber
        \end{equation}
        \State Form the conic representation using~\eqref{eq:conic}
        \State Solve the optimization problem using DR splitting with $u$ sub-iteration, as outlined in~\eqref{eq:DR_method}
        \State Update $\lambda_{t+1}$ using~\eqref{eq:lambdat1}
        \State Increment inner iteration index: $t = t + 1$.
    \Until{$\lvert I^{E}_{M} (\Omega^{t}_{s,m}, \tilde{a}_s^{i}) - \lambda_{t} P_{c} (\Omega^{t}_{s,m}, \tilde{a}_s^{i}) \rvert \leq \epsilon$}
    \State Increment outer iteration index: $i = i + 1$.
\Until{$\lvert \Gamma (\Omega^{t}_{s,m}, a_s^{i}) - \Gamma(\Omega^{t-1}_{s,m}, {a}_s^{i-1}) \rvert\! <\! \delta \Gamma (\Omega^{t-1}_{s,m}, {a}_s^{i-1})$}
\end{algorithmic}
\end{algorithm}
\setlength{\textfloatsep}{0.3cm}

\begin{figure*}[t]
\vspace{-0.5cm}
    \centering
    \begin{minipage}[t]{0.32\textwidth}
        \vspace{-4.7cm}
        \centering
        \includegraphics[trim=17 0cm 0cm 0cm,clip,width=1.1\textwidth]{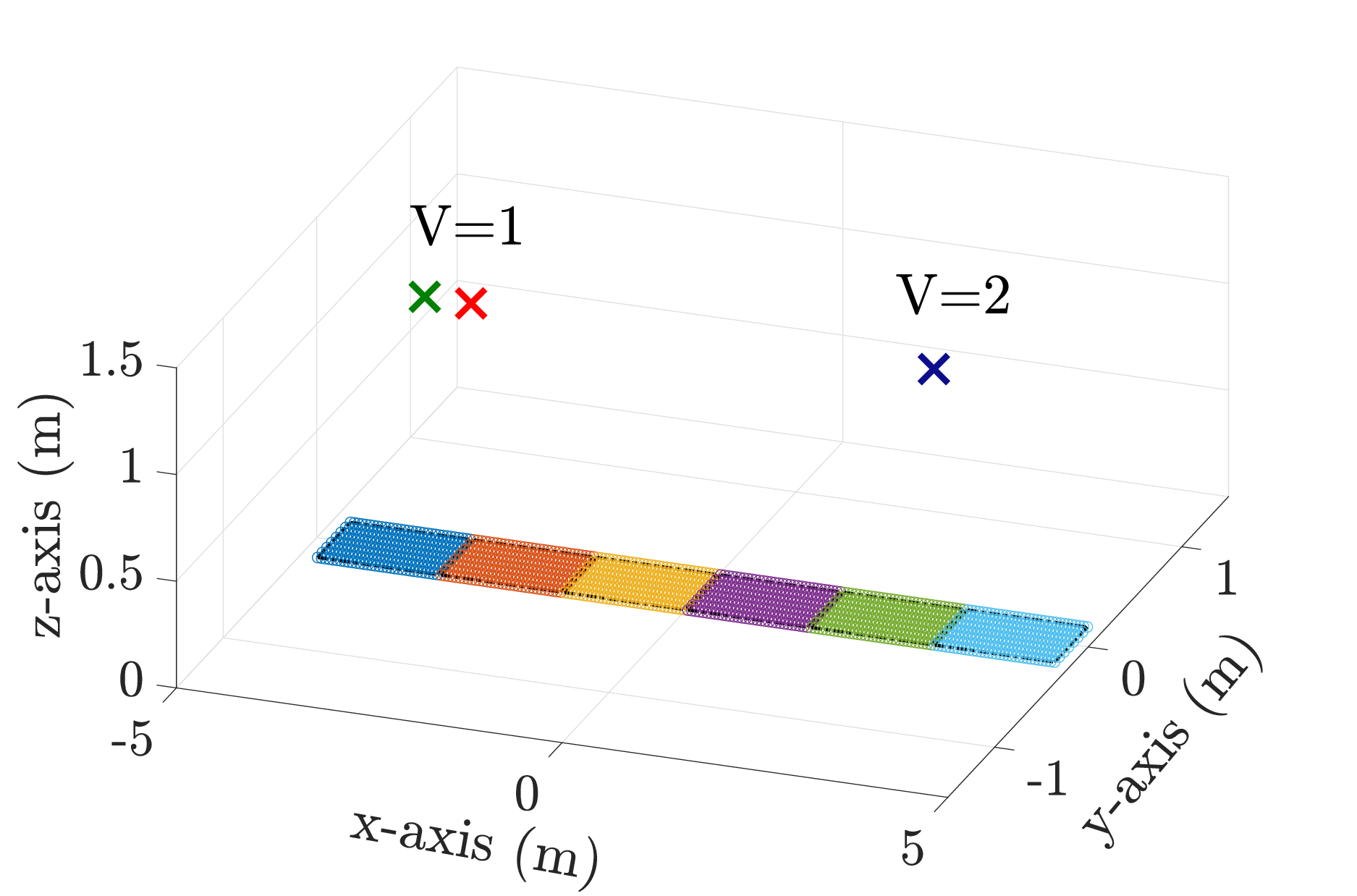}
        \vspace{-0.2em}
        \caption{\small XL-MIMO system with $6$ sub-arrays of $32 \times 8$ elements \& $3$ receivers in $2$ VRs.\normalsize}
    \label{fig:XL_MIMO}
    \end{minipage}
    \hfill
    \begin{minipage}[t]{0.32\textwidth}
        \centering
        \includegraphics[trim=0 0cm 0cm 0cm,clip,width=1.12\textwidth]{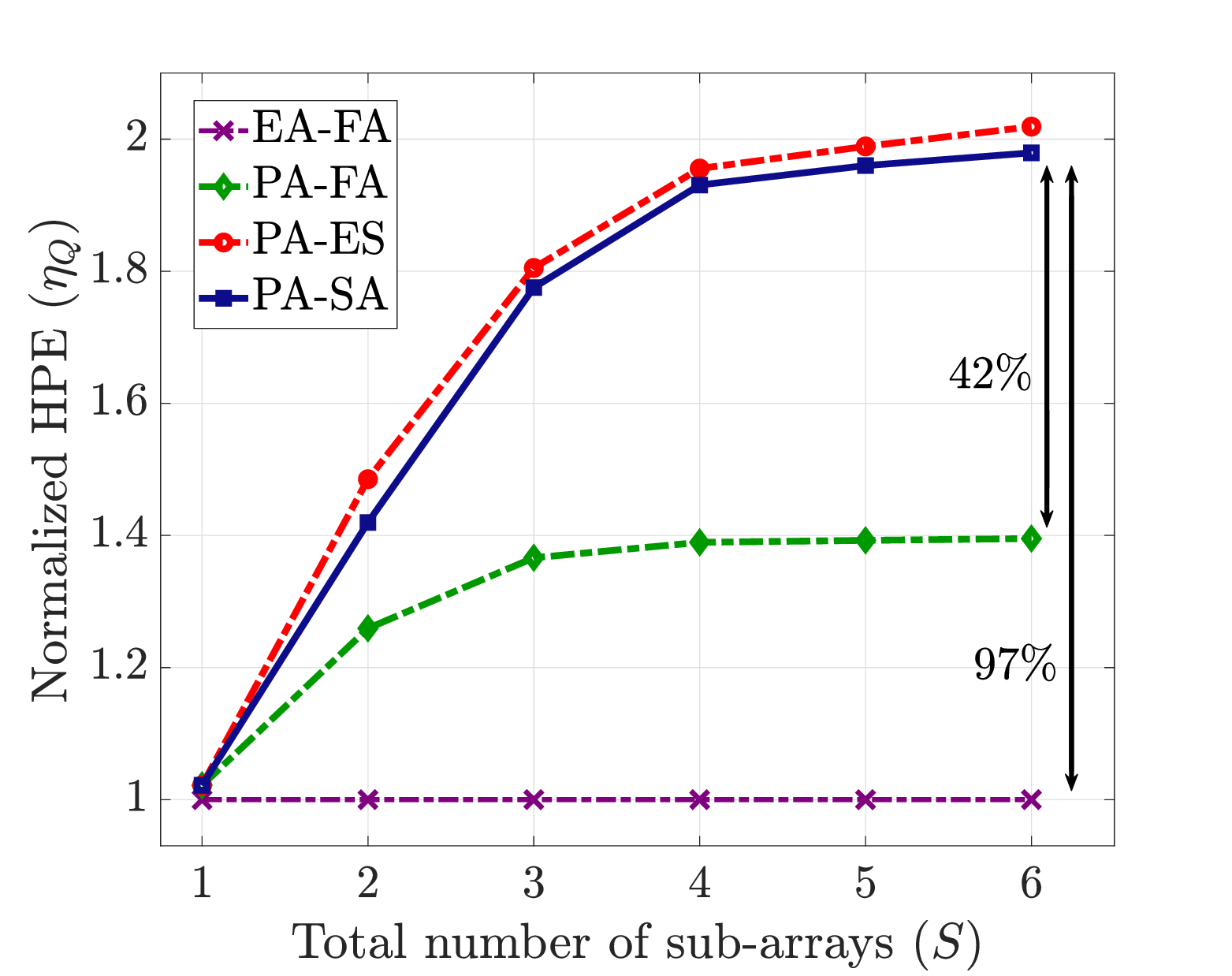}
        \vspace{-1.5em}
        \caption{\small Normalized HPE vs the number of sub-arrays for $V=1$.\normalsize}
        \label{fig:Energy_eff_V1}
    \end{minipage}
    \hfill
    \begin{minipage}[t]{0.32\textwidth}
        \centering
        \includegraphics[trim=0 0cm 0cm 0cm,clip,width=1.12\textwidth]{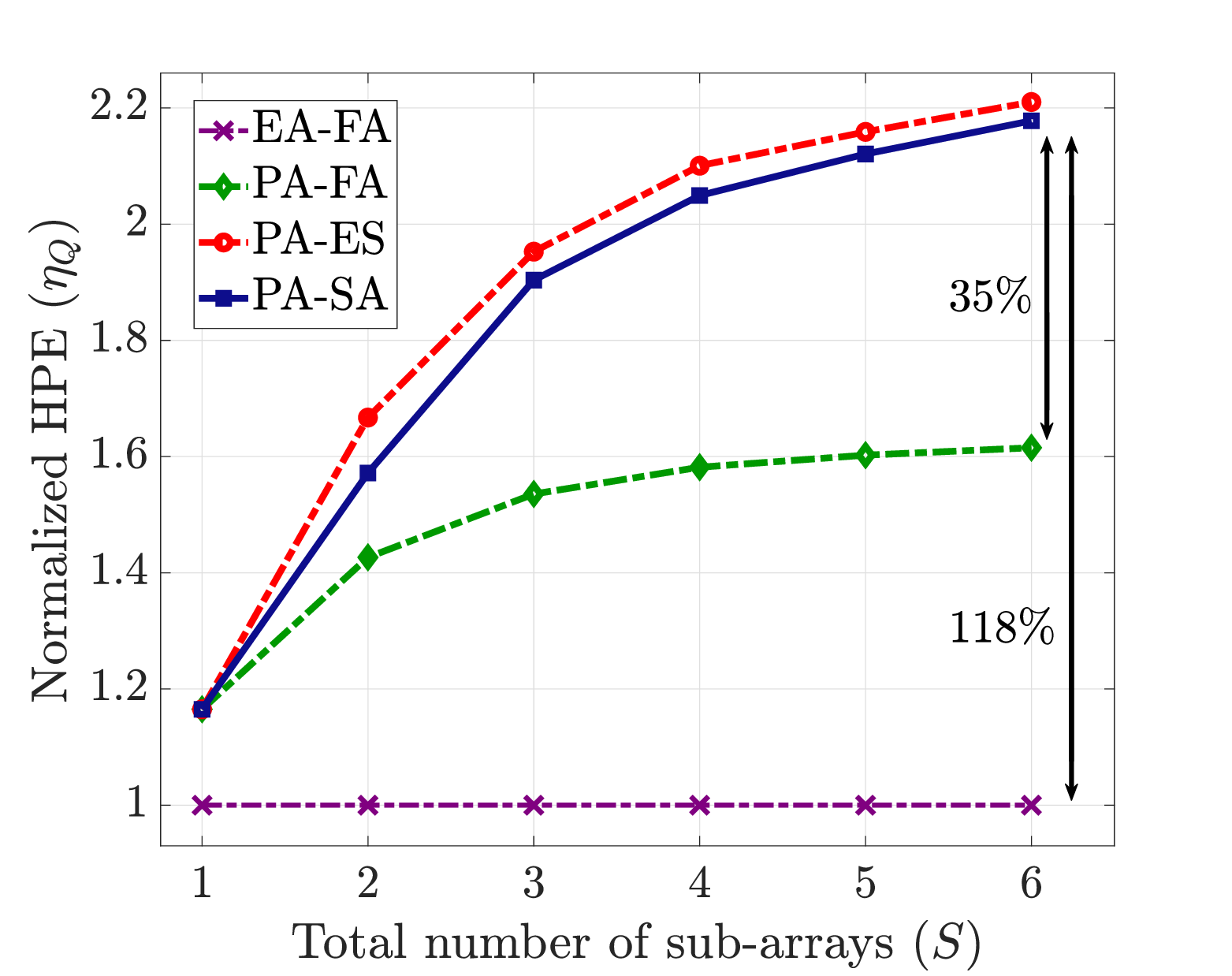}
        \vspace{-1.5em}
        \caption{\small Normalized HPE vs the number of sub-arrays for $V=2$.\normalsize}
        \label{fig:Energy_eff_V2}
    \end{minipage}
\vspace{-1.5em}
\end{figure*}

\vspace{-0.8em}
\subsection{Sub-array Activation Update (SA procedure)}
\label{subsec:subarray}
In this subsection, we explain the iterative SA procedure utilizing $\ots$ estimated by the PA procedure discussed in Section \ref{subsec:poweralloc}. For a specific iteration $i$ of the SA procedure, the PA procedure is completed for a particular SA $\as$ with the initial realization of the full array activation scenario $a^0_s =\{1\}^S$. Here, we introduce a surrogate auxiliary function $\gos$ which encompasses the normalized measure of the optimized PA estimates in iteration $(i-1)$, defined as   
\begin{align}~\label{eq:gios}
     \gios = \dfrac{\,\summ \osim}{\sus\summ \osim}.
\end{align}

Here, $\gos$ characterizes the ratio of the transmitted power by each sub-array relative to the total transmission power using the optimized PA from the PA procedure. This pertains to the fact that the sub-array which has higher $\gos$ value offers stronger contribution to the harvested power $\IEM$ with higher numerator values in the evaluation of the overall HPE in \eqref{eq:gios} in comparison to the dormant sub-arrays. Utilizing the function $\gos$ with a mean threshold $\ME\{\gios\}$, we can iteratively update the binary SA variables as given below:  
\begin{align}
\label{eq:binary_decision}
    \asi &= 
    \begin{cases}
        1 \quad \quad \gios \geq \ME\{\gios\},\\
        0 \quad \quad \gios < \ME\{\gios\}.
    \end{cases}
\end{align}

The choice of this baseline $\ME\{\gios\} (1/S)$ ensures that only those sub-arrays,  which have higher relative optimized PA coefficients than the mean value of the function $\gios$,  are activated in the next iteration. Thereby, it reduces the denominator value $\PC$ in \eqref{eq:gios}, leading to the HPE enhancement. At this stage, the updated discrete SA variables $\asi$ are scaled once again with the same continuous auxiliary function $\gios$ to define a parameterized SA variable $\tilde{a}_s^{i}$, which preserves the relative scaling within the activated sub-array subset during the next iteration of the PA procedure:
\begin{equation}\label{eq:parameterized}
     \tilde{a}_s^{i} = \gios \asi.
\end{equation}

The use of this continuous variable $\tilde{a}_s^{i}$ evaluated in \eqref{eq:parameterized} enables the elimination of less effective sub-arrays while monitoring the HPE improvement over iterations. Next, this SA choice $\tilde{a}_s^{i}$ is substituted in \eqref{eq:Dinkelbach} for invoking the next iteration $(t+1)$ of PA procedure. Although the users' association to a particular VR can be performed using the estimation techniques presented in the recent literature \cite{Tian}, we employ the optimized PA coefficients ($\os$) along with the binary decision-making SA variables ($\as$) to address the SnS effects of the channel between user $s$ and sub-array $m$.

\vspace{-1.1em}
\subsection{Overall Optimization Process and Complexity Analysis}
The PA procedure and SA procedure are executed alternatively until the point of near-optimality convergence with tolerance levels $\epsilon,\delta\! >\!0$ is achieved.\footnote{HPE is upper-bonded by the optimal PA-ES solution. The overall PA-SA algorithm converges to at least a locally finite optimal solution $\Gamma^*$. Mathematically, $\Gamma (\Omega^{t}_{s,m}, a_s^{i}) \geq \Gamma(\Omega^{t-1}_{s,m}, {a}_s^{i-1}), \, \Gamma (\Omega^{t}_{s,m}, a_s^{i}) \leq \Gamma^{*} < \infty $, using monotone convergence theorem: $\lim_{i \rightarrow \infty} \Gamma (\Omega^{t}_{s,m}, a_s^{i})  = \Gamma^{*}.$ } \footnote{Smaller levels of these convergence thresholds will achieve superior precision levels with higher computational needs and vice versa.} This process is summarized in \textbf{Algorithm~\ref{alg:opt_process}}. The implementation of the PA procedure has strong dependence on proximal operators within each DR iterations, with computational complexity  $\mathcal{O}(SM)$ for each sub-array-user pair. The number of iterations required for this sub-procedure convergence scales linearly with the problem size $\mathcal{O}(SM)$. Within the SA procedure, the surrogate function involves a calculation based on $SM$ power coefficients. Furthermore, the binary decision-making in \eqref{eq:binary_decision} over $\ME\{\gios\}$, leads to logarithmic scaling as $\mathcal{O}(\log(SM))$. Thus, the overall computational complexity of the proposed PA-SA algorithm is $\mathcal{O}(S^2M^2\log(SM))$. In comparison to the proposed SA procedure, the computational complexity of ES for optimal sub-array selection among all possible combinations is \big($\forall\,\as\!\in\!\{0,1\}^S$\big) is $\mathcal{O}(S^2M^2(2^S))$, which grows exponentially with $S$, i.e. number of sub-arrays. 

\vspace{-0.8em}
\section{Numerical Results}
In this section, we evaluate the performance of the proposed algorithm for the HPE optimization of a XL-MIMO system. This system is subdivided into $S$ sub-arrays, each equipped with $\ns\!=\nx \ny=\!256$ antenna elements which are arranged in $\nx\!=\!32$ and in $\ny\!=\!8$ elements with inter-antenna spacing $d\!=\!\lambda/2$. Each antenna element has the largest physical dimension as $D\!=\!\lambda/4$ with the carrier wavelength $\lambda\!=\!0.1$m \cite{Ramezani2024} and the near-field radiation boresight gain $b\!=\!2$ \cite{Haiyang}. We consider $M=3$ users which are located in $V=\{1,2\}$ distinct VRs, as shown in Fig.~\ref{fig:XL_MIMO}. The power consumption parameters  are set to the following values: $\varsigma=0.35$, $P_{et} = 50$ mW, $P_{\mathrm{syn}}=50$ mW, $P_{\mathrm{ct}}=48.2$mW and $P_{\mathrm{cr}}=62.5$ mW \cite{Jun_Zhang}.\footnote{Any increase in these parameters results in the degradation of $\gamEff$ performance, except $\varsigma$.}   The algorithm convergence thresholds are chosen as $\epsilon = 10^{-7}$ and $\delta =10^{-3}$ using an appropriate precision level of $0.1\%$ between the successive iterations.\footnote{The nominal value of $\IEM$ is $\sim 100 \mu$W, whose $0.1\%$ precision level relates to this particular $\epsilon$ choice.}

For comparison, four methods have been assessed for our HPE performance analysis: (i) \textbf{PA-SA}: proposed joint optimized PA-SA \big($\tilde{\Omega}_{s,m},\ats$\big) in \textbf{Algorithm \ref{alg:opt_process}} with optimized HPE without extensive computational needs for SA choice; (ii) \textbf{PA-FA}: PA procedure for optimized PA \big($\tilde{\Omega}_{s,m}$\big) with full array (FA) activation \big($\as\!=\!\{1\}^S$\big) with maximum power consumption; (iii) \textbf{PA-ES}: PA procedure for optimized PA \big($\tilde{\Omega}_{s,m}$\big) with ES using maximum computational needs; (iv)  \textbf{EA-FA}: equal PA \big($\os\!=\!P_s/M$\big) with FA \big($\as\!=\!\{1\}^S$\big). To demonstrate the relative gains, we consider the normalized HPE ratio, defined as $\eta_Q\triangleq\Gamma_{Q}\,/\,\Gamma_{\text{EA-FA}}$ for $Q\!\in\!\{\text{PA-SA, PA-FA, PA-ES, EA-FA}\}$, which represents the HPE of each method relative to the benchmark \textbf{EA-FA} with worst HPE and least computational requirements.

\begin{figure*}[t]
\vspace{-0.5cm}

    \centering
    \begin{minipage}[t]{0.24\textwidth}
        \centering
        \includegraphics[trim=0cm 0cm 0cm 0cm,clip,width=1.12\textwidth]{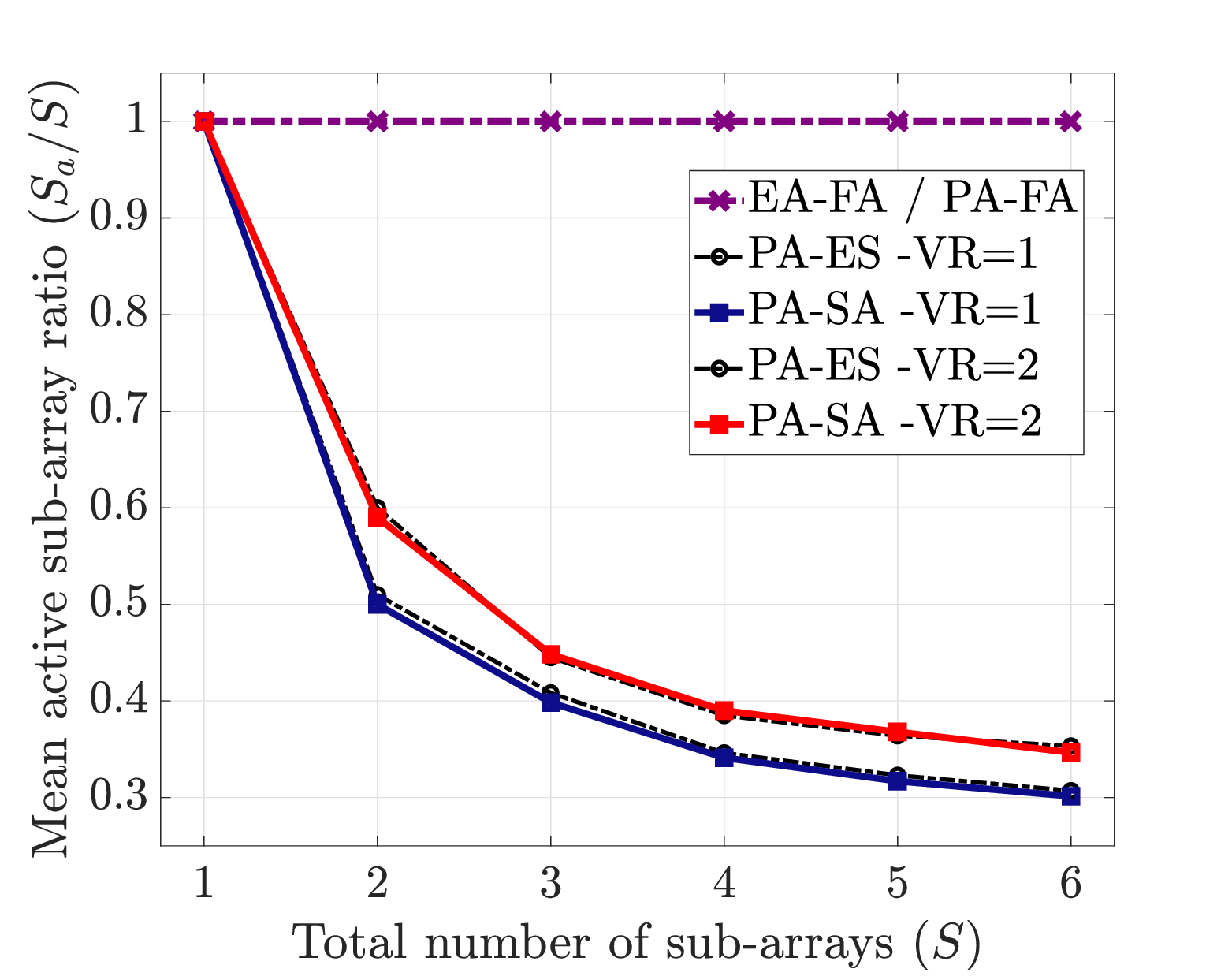}
        \vspace{-1.5em}
        \caption{\small Mean active sub-array ratio vs the number of sub-arrays.\normalsize}
        \label{fig:active_sub_array_ratio}
         \vspace{-1em}
    \end{minipage}
    \hfill
    \begin{minipage}[t]{0.24\textwidth}
        \centering
        \includegraphics[trim=0 0cm 0cm 0cm,clip,width=1.12\columnwidth]{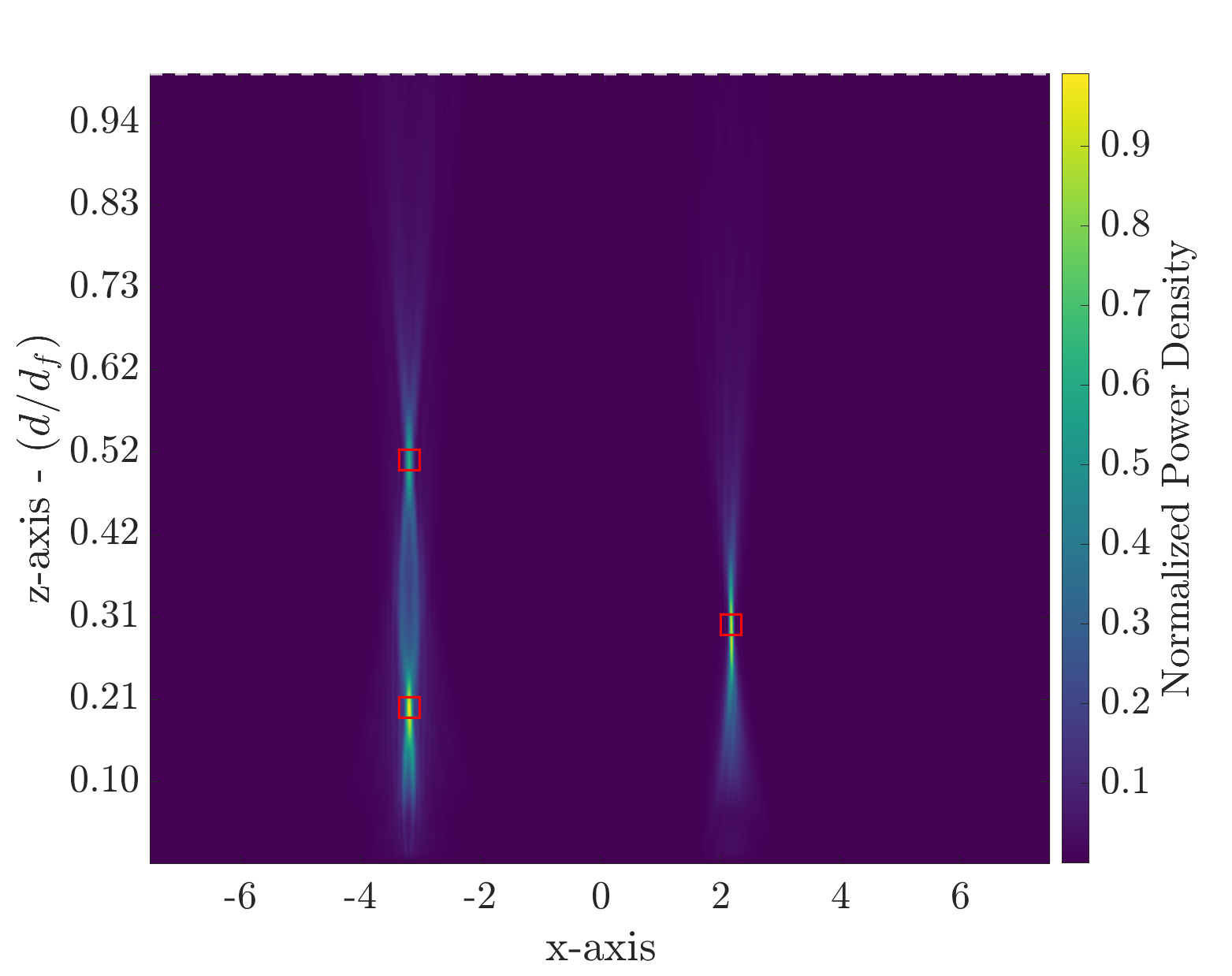}
         \vspace{-1.5em}
        \caption{\small Beamfocusing using near-field channel.\normalsize}
        \label{fig:Beamfocusing}
        \vspace{-1.5em}
    \end{minipage}
    \hfill
    \hspace{0.5em}
    \begin{minipage}[t]{0.24\textwidth}
        \centering
        \includegraphics[trim=0 0cm 0cm 0cm,clip,width=1.12\columnwidth]{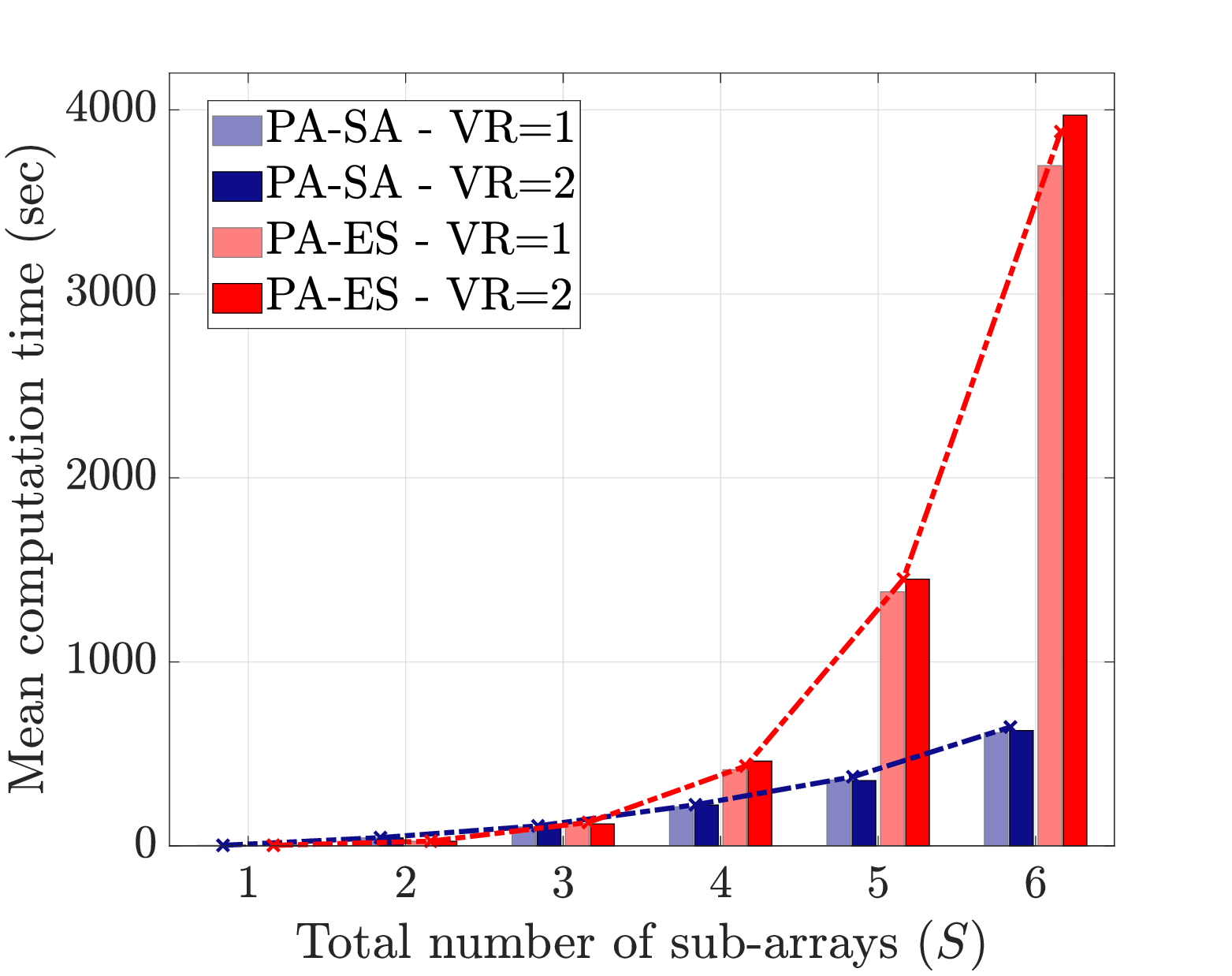}
         \vspace{-1.5em}
        \caption{\small Mean computation time vs the number of sub-arrays.\normalsize}
        \label{fig:Computation_time}
        \vspace{-1em}
    \end{minipage}
    \hfill
    \begin{minipage}[t]{0.24\textwidth}
        \centering
        \includegraphics[trim=0 0cm 0cm 0cm,clip,width=1.12\columnwidth]{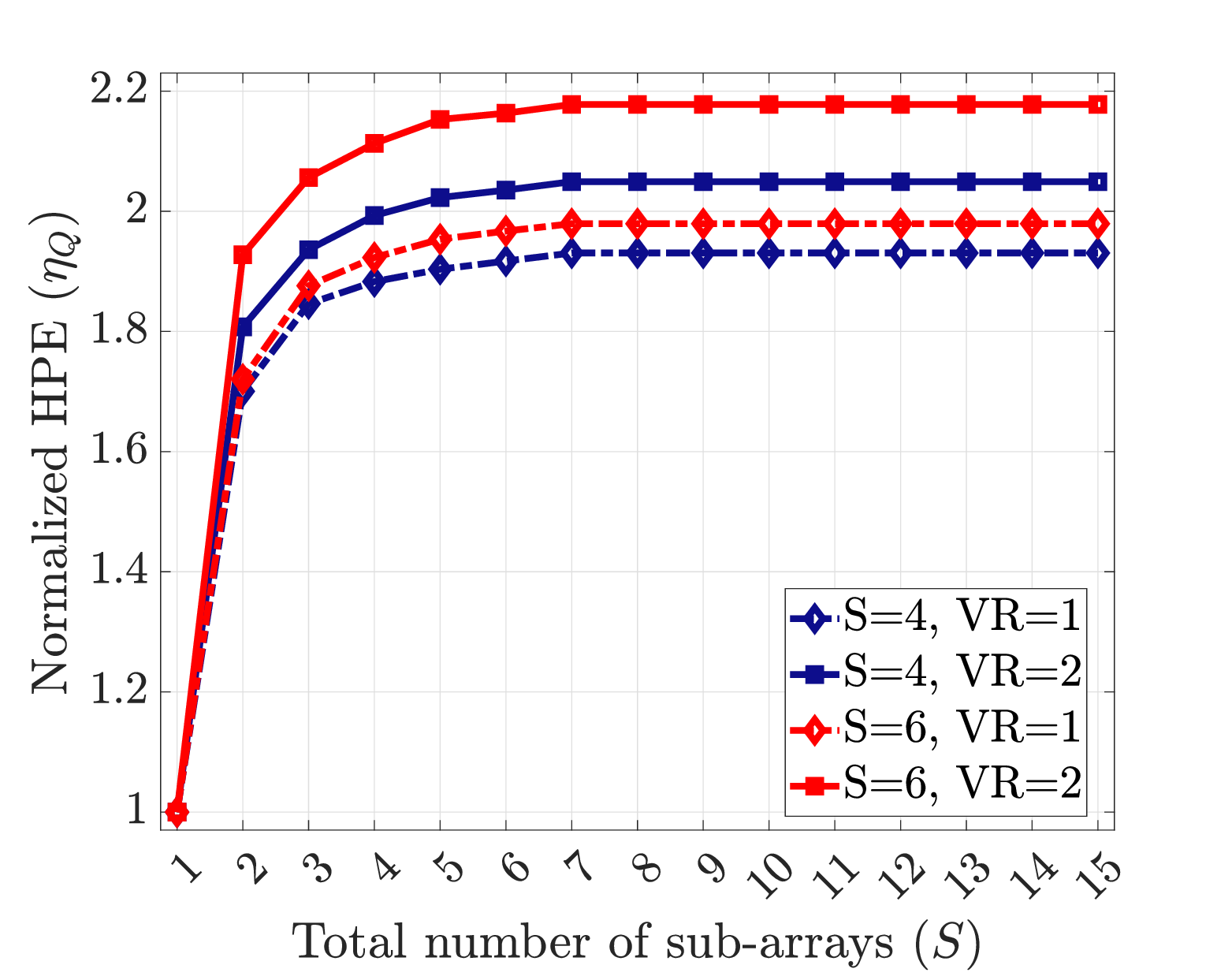}
         \vspace{-1.5em}
        \caption{\small Algorithm convergence vs the number of iterations.\normalsize}
        \label{fig:Convergence}
        \vspace{-1.5em}
    \end{minipage}
\vspace{-0.3em}
\end{figure*}

 This normalized HPE evaluation for different VR cases is presented in Fig. \ref{fig:Energy_eff_V1}\, for the $V=1$ case and in Fig. \ref{fig:Energy_eff_V2}\, for the $V=2$ case. It can be observed that the HPE performance of the SA choice, based on the proposed SA procedure (Section \ref{subsec:subarray}) along with PA procedure (Section \ref{subsec:poweralloc}), is very close to the ES for the optimal sub-array selection combined with the PA procedure. Moreover, considering the particular example of $S=6$ sub-arrays for the user positioning in $1$ \& $2$ VR regions, we can notice HPE performance gains of $97\%$ \& $118 \%$ for the joint optimization method respectively in comparison to the benchmark method of \textbf{EA-FA}. The HPE gains over \textbf{EA-FA} for the $V=2$ scenario are more significant than $V=1$, because the optimized power transfer from a specific sub-array subset will be more focused on the users co-located in a certain spatial region. On the other hand, the HPE gains over \textbf{PA-FA}  are observed as $42\%$ \& $35\%$ for $V=\{1,2\}$ VR scenarios. 

The mean ratio of the number of active sub-arrays ($S_a$) to the total number of sub-arrays ($S$) is another key parameter to be investigated in the HPE analysis. The results shown in Fig. \ref{fig:active_sub_array_ratio} reinforce the fact that a smaller proportion of sub-arrays is contributing to the HPE optimized-EH process with increasing array physical dimensions, as the users' VR is more confined to a certain segment of the XL-MIMO array. This ratio is lower in the single VR case ($30\%$) than in the multiple VR case ($35\%$), as fewer active sub-arrays suffice to serve the users' EH needs.

Position-dependent beamfocusing is a key feature of non-stationary near-field propagation, where spherical EM wavefronts can be manipulated to focus energy in both distance and angle. As shown in Fig. \ref{fig:Beamfocusing}, the received energy from the associated VR(s) is concentrated at user locations, even at the same angular depth, unlike traditional beamforming, which relies on angular EM beams. Furthermore, omitted results show that full-array and VR selection cases yield nearly equivalent performance.

In Fig. \ref{fig:Computation_time}, the mean computation time required for HPE optimization using the proposed algorithm (\textbf{PA-SA}) is compared with that of \textbf{PA-ES}. For these optimization simulations, the computational resources used are a CPU processor, 12th Gen Intel $\!\!$\textsuperscript{\tiny{\textregistered}} $\!\!$ Core $\!\!\!\!$ \textsuperscript{\tiny{\texttrademark}}  i7-1270P (average clock speed 3.50 GHz) with 16GB RAM. It is evident that our proposed algorithm clearly outperforms the \textbf{PA-ES} method, as the mean computational time for the proposed \textbf{PA-SA} method increases linearly with an increasing number of sub-arrays ($S$), while it increases exponentially for the \textbf{PA-ES} case. 

The convergence analysis of the proposed \textbf{PA-SA} algorithm is presented in Fig. \ref{fig:Convergence} for different VR ($V\!\in\!\{1,2\}$) and sub-array ($S\!\in\!\{4,6\}$) scenarios. We can observe that the proposed algorithm can rapidly converge to the near-optimal solutions within the first $10$ iterations, while improving the HPE performance monotonically. It can be further seen that within the first $2$ iterations, the HPE performance reaches at average $88\%$ of the final convergence level. Moreover, the HPE metrics for both sub-array setups ($S\!\in\!\{4,6\}$) for the case $V=2$ outperform those for the case $V=1$, since the users are tightly co-located in more confined VRs at smaller distances to the corresponding sub-arrays.

\vspace{-0.5em}
\section{Conclusion}
In this letter, we have developed an optimization framework to enhance the HPE of an XL-MIMO system in near-field communication, addressing challenges such as spatial non-stationarities. The proposed joint optimization algorithm consolidates the PA and SA using useful techniques, such as the Dinkelbach's transform based fractional programming and the DR splitting method. The simulation results show that our proposed method achieves near-optimal sub-array selection, while significantly reducing the computational requirements compared to the ES method. Future research could investigate XL-MIMO for SWIPT using appropriate channel estimation along with more energy-efficient hybrid beamforming techniques. Joint optimization of analog phase shifters and digital precoding can reduce the required number of RF chains, thereby mitigating the high power consumption associated with fully digital precoding.

\vspace{-1.0em}
\bibliographystyle{IEEEtran}
\bibliography{main}

\end{document}